\documentstyle[iopconf1,epsf,epsfig]{article}

\def\ga{\mathrel{\raise.3ex\hbox{$>$\kern-.75em\lower1ex\hbox{$\sim$}}}}
\def\la{\mathrel{\raise.3ex\hbox{$<$\kern-.75em\lower1ex\hbox{$\sim$}}}}
\def\gev{{\rm \, Ge\kern-0.090em V}}
\def\tev{{\rm \, Te\kern-0.090em V}}

\def\ss{\scriptscriptstyle}
\def\msf{m_{\tilde f}}
\def\ohsq{\Omega_{\tilde\chi} h^2}
\def\m12{m_{1\!/2}}

\def\msn{m_{\tilde\nu}}

\def\bino{\widetilde B}
\def\neut{\tilde\chi^0}

\def\mchi{m_{\chi}}
\def\mcha{m_{\chi^{\pm}}}
\def\cha{\tilde\chi^{\pm}}
\def\ohsq{\Omega_{\chi} h^2}
\def\tb{\tan\beta}
\begin{document}      
\rightline{MADPH-98-1071}

\title{LEP Constraints on Neutralino Relic Densities, and the 
Fate of Higgsino Dark Matter}

\author{Toby Falk}
\affil{Department of Physics, University of Wisconsin, Madison, WI 53706, USA}

\beginabstract We examine the current state of neutralino dark matter
and consider how the LEP constraints on the Minimal Supersymmetric
Standard Model parameters are squeezing the available dark matter
regions.  We also show how cosmological constraints augment bounds
coming from collider searches to further constrain the MSSM parameter
space.  \endabstract

\section{Introduction}

If $R$-parity is conserved, the Lightest Supersymmetric Particle (LSP)
is stable.  In many supersymmetric models, the LSP tends to have a
cosmologically interesting relic density and is a good dark
matter candidate \cite{gold,susydm1}, and it is a very appealing feature
of the Minimal Supersymmetric Standard Model that it can naturally
provide an answer to the dark matter question.  However, due the
characteristically large relic abundance of the LSP's, the MSSM is also naturally susceptible
to cosmological constraints, either because the relic density of LSP's is so
large that it is in conflict with a gross cosmological feature,
e.g.  the age of the universe, or because one may be able to detect
the interactions of the LSP with terrestrial detectors or detect the
self-interactions of the LSP in the galactic halo or in the cores of
the earth or the sun \cite{det}.

The MSSM contains in principle two neutral dark matter candidates, the
sneutrino $\tilde\nu$, and the lightest neutralino $\neut$.  The
sneutrino, though, has already been excluded\footnote{Outside the
  MSSM, these constraints may be evaded.  See \cite{hmm},\cite{hh}.}
as a dark matter candidate by a combination of LEP bounds and direct
detection experiments \cite{fos1} .  The lightest neutralino may be
either gaugino-like (in particular $\bino$-like) or Higgsino-like, and
both the phenomenology and cosmology of neutralinos depend strongly on
the neutralino composition.  In this talk we will study the question of how
viable these remaining MSSM dark matter candidates are in light of recent LEP data.

\section{Neutralinos}
\begin{figure}[thb]
\begin{center}
\vspace*{-1.2in} 
\hspace*{-0.2in}
\epsfig{file=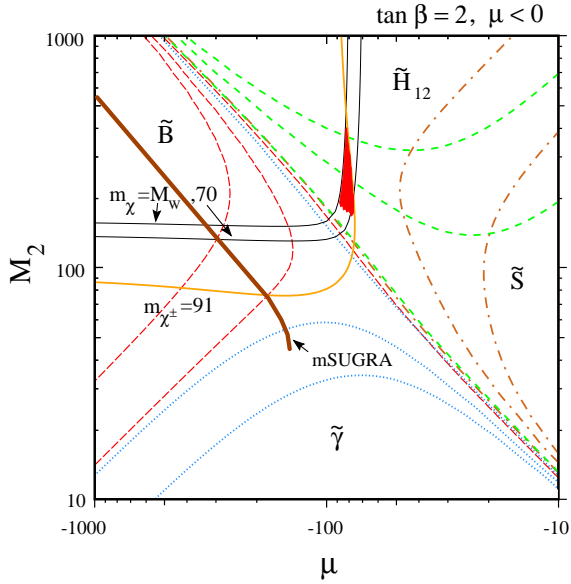,height=5in}
\end{center}
\vspace{-1.0in} 
\caption{{\it Contours of neutralino purity: 99\%, 97\% and 75\%, and chargino and neutralino masses (solid lines). The
    long-dashed lines are contours of high bino purity, the dotted
    lines are contours of high photino purity, the dashed lines are
    contours of high ${\tilde H}_{12}$ Higgsino purity, and the
    dash-dotted lines are contours of high ${\tilde S}_0$ Higgsino
    purity. Also shown are contours of constant $\mchi$ and $\mcha$
    and the dependence of $\mu$ on $M_2$ in mSUGRA. We've taken 
    $m_0=100\gev$. }} {
  \label{fig:purity} } \vspace{0.3in}
\end{figure}

In general, the  neutralinos are  linear combinations of the neutral
gauginos and Higgsinos,
\begin{equation}
\chi_i = \beta_i  {\tilde B} + \alpha _i{\tilde W} + \gamma_i{\tilde
H}_1 + \delta _i{\tilde H}_2, \hspace{0.3in}  i=1,\ldots, 4
\label{neut}
\end{equation}
In this notation, the gaugino purity of a neutralino
$\chi_i$ is defined to be $\sqrt{{\alpha_i}^2 + {\beta_i}^2}$, and its Higgsino
purity $\sqrt{{\gamma_i}^2 + {\delta_i}^2}$.  In the $({\tilde B}, {\tilde W}^3, {{\tilde H}^0}_1,{{\tilde
    H}^0}_2 )$ basis, the neutralino mass
matrix takes the form
\begin{equation}
\left( \begin{array}{cccc}
M_1 & 0 & {-M_Z s_\theta c_\beta} &  {M_Z s_\theta s_\beta} \\ 0 & 
M_2 & {M_Z c_\theta c_\beta} & {-M_Z c_\theta
s_\beta}
\\ {-M_Z s_\theta c_\beta} & {M_Z c_\theta c_\beta} & 0
& -\mu
\\ {M_Z s_\theta s_\beta} & {-M_Z c_\theta s_\beta} &
-\mu & 0 
\end{array} \right) ,
\label{mm}
\vspace{0.1cm}
\end{equation}
where $s_\theta$ ($c_\theta$) $=\sin\theta_W$ ($\cos\theta_W$),
$s_\beta$ ($c_\beta$) $=\sin\beta$ ($\cos\beta$), and where gaugino
mass unification implies $M_1= 5/3 \tan^2\theta_W M_2$.  The
coefficients $\alpha_i,\ldots, \delta_i$ in (\ref{neut}) 
depend on $\tb$ and on the
soft SUSY-breaking mass parameters $M_1, M_2$ and $\mu$, which appear
in (\ref{mm}), and Fig.~1 displays the regions of high gaugino and Higgsino
purity in the $(\mu,M_2)$ plane.  In the limit
$|\mu|\gg M_i$, the lightest neutralino is gaugino-like, specifically
a $\bino$, i.e. $\beta_1\approx 1$ in ($\ref{neut}$), and this is
typically the case in mSUGRA and in models with gauge mediated
supersymmetry breaking (GMSB).  In Fig.~\ref{fig:purity}, contours
of constant $\bino$ purity are displayed as
long-dashed lines.  In the opposite limit, $M_i\gg|\mu|$, the lightest
neutralino is Higgsino-like, and for small to moderate $M_2$, the
lightest neutralino is the particular Higgsino combination defined by ${\tilde
  S}^0 \equiv {\tilde H}_1^0 \cos \beta + {\tilde H}_2^0 \sin \beta$,
i.e., $\gamma = \cos \beta$ and $\delta = \sin \beta $, with
$m_{\tilde S_0} \rightarrow \mu \sin 2 \beta$~\cite{EHNOS}. Contours
of ${\tilde S}^0$ purity are displayed in Fig.~\ref{fig:purity} as
dash-dotted lines.  For large $M_2$, the lightest neutralino is the
state ${\tilde H}_{12} \equiv {1 \over \sqrt{2}} ({\tilde H}_1^0 \pm
{\tilde H}_2^0)$, i.e., $\delta = \pm \gamma = \pm 1/\sqrt{2}$ for
sgn$(\mu) = \pm 1$, with $m_{\tilde H_{12}} \rightarrow
|\mu|$~\cite{osi3}, and contours of ${\tilde H}_{12}$ purity are shown as
short-dashed lines in Fig.~\ref{fig:purity}.

In mSUGRA, the relationship between $\mu$ and $M_2$ is determined from
the conditions of gaugino unification and sfermion and Higgs mass
unification, along with the requirement of correct radiative
electroweak symmetry breaking, and this relationship is shown in Fig.~\ref{fig:purity} 
as a thick solid line.
As suggested above, the resulting contour lies in the
gaugino region, and the heavier the neutralino, the more pure its
$\bino$ content.  Also shown in Fig.~\ref{fig:purity} are contours of
constant chargino mass, where $91\gev$ represents the current LEP
lower bound on $\mcha$ \cite{LEPC}, and constant neutralino mass.  As
a preview of the small size of the Higgsino dark matter region we'll
be discussing, it consists of a subset of the shaded region between
the $\mcha=91\gev \mbox{ and } \mchi=M_W$ contours.

\vspace{-0.5in}
\section{LEP Bounds}
\label{sec:lepbounds}

Recent runs at LEP at center-of-mass energies of 172 and 183 $\gev$
have excluded large areas of MSSM parameter space, and subsequent runs
at $\sim 190$ and $200\gev$ will push the bounds even further.  In
this section we summarize the LEP bounds that we implement in this
talk. 
\begin{itemize}
\item Searches for chargino pair production at LEP 172 and LEP 183
  have excluded chargino masses less than $86\gev$ \cite{cha172} and
  $91\gev$\cite{LEPC} respectively, modulo two loopholes.  The first
  loophole is relevant when the neutralino is a gaugino and occurs
  when the mass of the sneutrino is close to that of the chargino.
  The lower limit on $\mcha$ is reduced as $\msn$ is reduced toward
  $\mcha$ from above, due to destructive interference with the
  t-channel sneutrino exchange process, and then disappears entirely
  for $\mcha > \msn \ga \mcha - 3\gev$, in which case $\cha$ decay is
  dominated by $\tilde\nu +$ soft lepton final states.  Bounds on
  chargino production reappear when $\mcha-\msn\ga 3\gev$ and the
  lepton detection efficiency picks up again.  The second loophole is
  relevant in the Higgsino region and occurs when $\Delta
  M=\mcha-\mchi$ is small, and it again is due to the reduction in
  detection efficiency when the mass difference between the produced
  particle and its supersymmetric decay product is small.  The effect
  of each of these loopholes will be discussed in more detail later
  on.
  
\item Searches for associated neutralino production can provide
  strong bounds when the neutralino is a Higgsino.  In the
  Higgsino region of interest for this talk, associated neutralino
  production is dominated by $e^+e^- \rightarrow \chi \chi_{2,3,4}$
  and corresponds essentially to $m_{\chi} + m_{\chi'_H} = 182\gev$,
  where $\chi'_H$ is the lightest mainly-Higgsino state among the
  $\chi_{2,3,4}$.
  
\item Searches for Higgs production provide a lower limit of
  $m_h>88\gev$ at low $\tb$ \cite{higgs}.  Bounds from Higgs searches
  are particularly constraining at low $\tb$, where the experimental
  bounds are strongest and where the tree level Higgs mass is small.
  Here radiative corrections to the Higgs mass \cite{MSSMHiggs} must
  be very large, leading to strong lower bounds on the masses of the
  sfermions, and in particular the stops.  However, the extraction of
  the radiatively corrected Higgs mass in the MSSM has an uncertainty
  of $\sim2\gev$, so we conservatively take $m_h>86\gev$ as our
  experimental lower limit at low $\tb$.
  
\item We also implement bounds coming from searches for sfermion
  production, in particular slepton \cite{alephslps} and stop
  production \cite{stopLEP}, as well as constraints on the sneutrino
  mass from the $Z$ width \cite{efos}.

\end{itemize}

\section{Gaugino Dark Matter}
  
As we have seen above, the lightest neutralino tends in many instances
to be a $\bino$, as is the case in mSUGRA, in particular.  For our
numerical examples, we will restrict our attention to mSUGRA, although
the qualitative features apply to more general cases, so long as the
LSP is a $\bino$.  The relic abundance of neutralinos is given by
\begin{equation}
\ohsq\sim{3\times10^{-10}\gev^{-2}\over\langle\sigma_{\rm
    ann}v\rangle} < 0.3,
\label{ohsq}
\end{equation}
where $\langle\sigma_{\rm ann}v\rangle$ is the thermally averaged
neutralino annihilation cross-section at the time the neutralinos fall
out of chemical equilibrium with the thermal bath.  The inequality in
(\ref{ohsq}) comes from the requirements that the age of the universe
$t_U$ be greater than 12 Gyr and that the total $\Omega<1$.  

\begin{figure}[thb]
\begin{center}
\epsfig{file=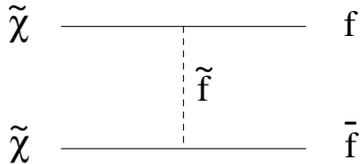,height=0.8in}
\end{center}
\caption{{\it The dominant contribution to $\bino$ annihilation.}}
{ \label{fig:feyn} }
\end{figure}

In the early universe, $\bino$ annihilation is typically dominated by
sfermion exchange into fermion pairs (Fig.~(\ref{fig:feyn})).  Due to
the Majorana nature of the neutralinos, this process generally
exhibits a p-wave suppression \cite{gold} (but see \cite{fmos,fos})
which decreases $\langle\sigma_{\rm ann}v\rangle$, and increases
$\ohsq$, by roughly an order of magnitude.  Now, as the sfermion
masses are increased, the relic abundance of neutralinos increases,
and for sufficiently heavy sfermions, $\ohsq$ violates the bound
(\ref{ohsq}).  Thus the requirement that $t_U>12$ Gyr translates into
an upper bound on the sfermion masses.

In mSUGRA, the sfermion and gaugino masses are given by
\begin{eqnarray}
\msf^2 &=& m_0^2 + C_{\tilde f}\, \m12^2 + O(m_Z^2), \\
(M_1, M_2)&\approx& (0.4, 0.8)\; \m12,
\end{eqnarray}
where $m_0$ and $\m12$ are the common scalar and gaugino mass
parameters, and the coefficients $C_{\tilde f}$ are determined by the
RGE evolution of the sfermion masses from the unification scale to the
electroweak scale.  The remaining parameters in mSUGRA are the common
trilinear scalar mass parameter $A_0$, which appears in the sfermion
mass matrices, $\tb$, and the sign of $\mu$.  The magnitude of $\mu$
is fixed in mSUGRA by the conditions of correct electroweak symmetry
breaking, as seen in Fig.~\ref{fig:purity}.  The sfermion masses are
typically insensitive to $A_0$, so in mSUGRA, the bound on $t_U$
translates simply into a upper bound on $m_0$ and $\m12$.  Because the
cosmological constraint provides upper limits on the soft masses,
whereas particle searches typically give lower limits on the same
parameters, the two types of constraints are nicely complementary.
Lastly, the Higgs mass is given by
\begin{equation}
m_h^2=m_Z^2 \cos^22\beta \;+\; {\rm rad.\;corr.\,} (m_{\tilde t_i}^2,m_t,A_t),
\end{equation}
where the dominant part of the radiative corrections depends
logarithmically on the stop masses.  At low $\tb$, the tree level
Higgs mass is small, and the Higgs mass
constraint then  imposes severe lower bounds on the stop masses, which
translates into lower bounds on $\m12$ and $m_0$.

\begin{figure}[thb]
\begin{center}
\epsfig{file=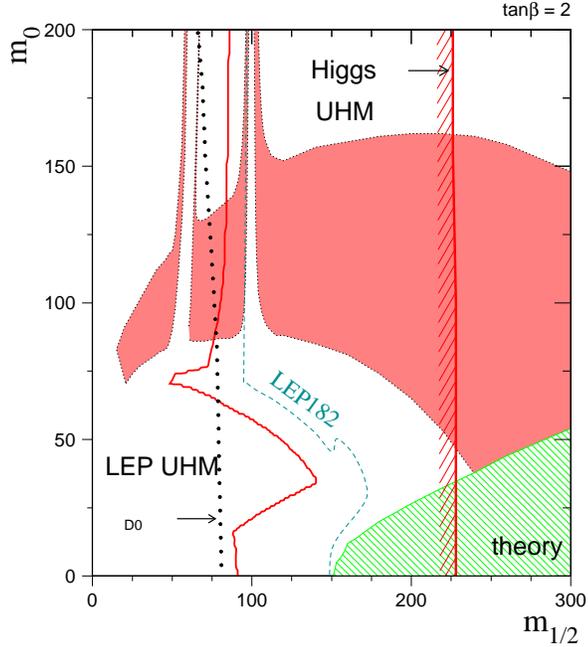,height=3.8in}
\end{center}
\vspace{-0.6in} 
\caption{{\it 
    We display for $\mu < 0$ and $\tan\beta=2 $ the domains of the
    ($m_{1/2}$,~$m_0$) plane that are excluded by the LEP 172 chargino
    and selectron searches in mSUGRA, the domains that are excluded by
    Higgs searches, the regions that are excluded cosmologically
    because $m_{{\tilde \tau}_R} < \mchi$, and the domains that
    have relic neutralino densities in the favoured range
    $0.1<\ohsq<0.3$}} 
{ \label{fig:msugra} } 
\end{figure}

Fig.~\ref{fig:msugra} summarizes \cite{efos2} the LEP 172 constraints
on the mSUGRA parameter space for $\tb=2,\, \mu<0$.  The solid contour
at the left of the figure represents the combined chargino and slepton
bounds\footnote{All the curves in Fig.~\ref{fig:msugra} are computed
  using tree level neutralino and chargino masses.  They shift by a
  few $\gev$ when 1-loop radiative corrections to the masses are
  included.}.  At large $m_0$, the chargino bound reaches the kinematic
limit, but as $m_0$ is decreased, the sneutrino mass falls, and the
first loophole described in section \ref{sec:lepbounds} reduces the
bound below $86 \gev$.  At $m_0\sim 75\gev$, the sneutrino mass drops
below that of the chargino, and the bound retreats rapidly to the
left; at lower $m_0$, the slepton bound comes into play and shuts the
door to lower $\m12$.  At this relatively low value of $\tb$, 
both the chargino and slepton bounds are dominated by the Higgs mass
constraint ($\sim 76\gev$), shown as the thick solid line stretching
almost vertically at $\m12\sim250 \gev$.  For lower values of $\tb$,
the Higgs bound moves rapidly to the right, whereas for higher $\tb$, the
chargino bound moves to the right, and the Higgs bound retreats to the
left and crosses beneath the combined chargino/slepton bound at
$\tb\sim3$.

\begin{figure}[thb]
\begin{center}
\vspace*{-0.6in} 
\epsfig{file=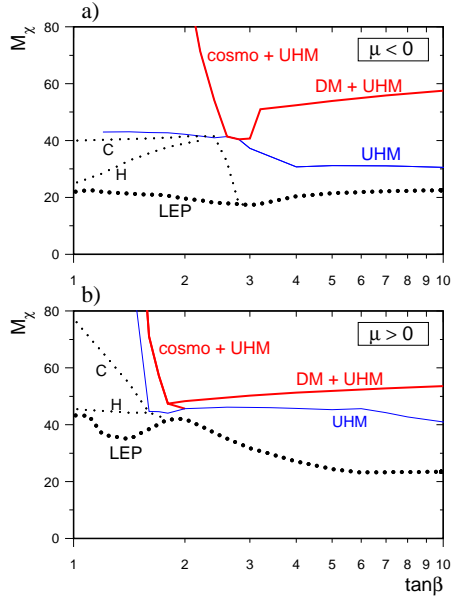,height=4.2in}
\end{center}
\vspace{-0.8in} 
\caption{{\it Various lower limits on $m_{\chi}$ using different
    experimental and theoretical inputs are compared, as functions of
    $\tan\beta$, for both (a) $|\mu| < 0$ and (b) $|\mu| > 0$.  The
    solid curves represent bounds in mSUGRA, including purely
    experimental bounds (``UHM''), separately augmented by cosmological
    (``cosmo+UHM'') and dark matter (``DM+UHM'') constraints. The
    dotted curves show the impact of Higgs (H) and cosmological (C)
    constraints in the MSSM.}} {
  \label{fig:mchimin} } \vspace{0.1in}
\end{figure}

It is interesting to consider how low a neutralino mass these combined
constraints admit.  The Higgs bound forces one to very large $\m12$ at
low $m_0$, giving a heavy neutralino; however, one can make the stops
heavy by taking $m_0$ very large, rather than $\m12$, and so the Higgs
contour does bend to the left at very large $m_0$ and strikes the
chargino contour.  At $\tb=2$, this occurs at $m_0\sim800\gev$.
Consequently, in the absence of an independent upper limit on $m_0$,
the lower bound on $\mchi$ is set by the intersection of the Higgs
contour with the chargino contour at large $m_0$ and yields a minimum
neutralino mass of $42\gev$.  Fig.~\ref{fig:mchimin} displays
\cite{efos2} the lower bound on $\mchi$ as a function of $\tb$, given
different sets of theoretical and experimental constraints.  The
purely experimental lower bound (i.e. not yet using cosmology) in
mSUGRA is given by the solid line labelled UHM (as a reminder that it
is the Unification of the Higgs Masses with the sfermion masses which
leads to $|\mu|$ being an output, rather than an input, in mSUGRA).

The dark shaded area in Fig.~\ref{fig:msugra} delimits the
cosmologically preferred region with $0.1<\ohsq<0.3$.  The upper
limit, as described above, comes from an upper bound on the age of the
universe.  The lower limit is less a bound {\it per se} than a
preference, stemming from the desire to have the neutralinos comprise
a significant fraction of the dark matter.  The two narrow vertical
channels at $\m12\sim65\gev$ and $\m12\sim100\gev$ arise from
s-channel neutralino annihilation on the Higgs and $Z^0$ poles,
respectively.  Note that top of the shaded area continues to fall for
$\m12>300\gev$, and it intersects with the ``theory'' excluded area
(where a stau is the LSP) at $\m12\sim425\gev$.  Thus, as advertised,
$\ohsq<0.3$ yields an upper bound on both $m_0$ and $\m12$.  This
independent upper bound on $m_0$ forbids the large $m_0$ solution to
the Higgs mass constraint (modulo a very narrow region on the $Z^0$
pole) and dramatically increases the lower bound on the neutralino
mass for values of $\tb$ where the Higgs mass bound is significant.
This is seen in Fig.~\ref{fig:mchimin}, where the branch of the solid
curve labelled ``cosmo+UHM'' is the result of including both the
constraints from particle searches at LEP~172 and the requirement
$\ohsq<0.3$.

Further, for low enough $\tan\beta$, the Higgs bound moves entirely to
the right of the dark-shaded region, and for $\mu<0$, the
cosmologically allowed range\footnote{The cosmological upper bound on
  $\m12$ varies only weakly with tan$\beta$ for tan$\beta \la 2$. } with
$\ohsq<0.3$ is actually {\em incompatible} with the Higgs lower limit
on $m_{1/2}$ for tan$\beta \la 1.7$. We conclude that in mSUGRA there
is no range of $m_{1/2}$ compatible with all the constraints provided
by the LEP 172 particle searches and the upper bound on the
cosmological relic density, for sufficiently small tan$\beta \la 1.7$.
Hence, {\em there is a lower bound} $\tb\ga 1.7$, if all these
constraints are applied simultaneously.  Similarly, for $\mu > 0$, the
bound from the LEP 172 searches is $\tan\beta~\ga~1.4$.  These bounds
have since improved.  Updated chargino and slepton contours from LEP
183 are shown as dashed lines in Fig.~\ref{fig:msugra}.  The chargino
contour displayed is the kinematic limit of $91\gev$, and as discussed above,
overestimates the chargino bound for low $m_0$.  The LEP 183 Higgs
bound is off the right side of the figure and in fact is very close to
the intersection point of the $\ohsq<0.3$ and ``theory'' contours.
The LEP 183 Higgs bound then combines with the cosmological constraint
to yield the updated \cite{efgos} lower bounds
\begin{equation}
\tb\ga\left\{
  \begin{array}{@{\hspace{0.3cm}}l@{\hspace{0.7in}}r}
    2.0 & \mu<0 \\ 
    1.65&\mu>0
\end{array}\right.
\end{equation}
  
Lastly, the branch of the solid curve labelled ``DM+UHM'' in
Fig.~\ref{fig:mchimin} exhibits the effect of additionally including
the constraint that there be a significant amount of neutralino dark
matter, $\ohsq>0.1$.  As $\tb$ is increased, the chargino bound moves
to the right, as does the Higgs pole, and the Higgs and $Z_0$ poles
widen, eventually merging to form one large pole region, at
$\tb\sim3$.  The kink in the ``DM+UHM'' line emerges from the
necessity to sit on the right side of the combined pole region, after
the merging, in order to have sufficient dark matter.

\section{Higgsino Dark Matter}
\subsection{Introduction}

As shown in Fig.~\ref{fig:purity}, when the neutralino is a Higgsino,
it approaches the particular combination ${\tilde H}_{12}$.  There are
in principle two regions in the ($\mu, M_2$) plane where ${\tilde
  H}_{12}$ can provide an interesting relic density.  For large $M_2$,
they correspond to $|\mu| < M_W$ but with $m_{\chi^{\pm}}$ above the
LEP limit, and $|\mu| \ga 1$ TeV~\cite{osi3,ge}.  The
intermediate-mass Higgsino states are not of cosmological interest,
because of their rapid annihilations to $W$ and $Z$ pairs. We have
little to add concerning the very heavy ${\tilde H}_{12}$ states, but
the lighter Higgsinos lie directly in the region where the current LEP
runs are eating away at the parameter plane, and thus it is of
interest to examine how much of the light Higgsino parameter space
remains consistent with the LEP bounds.

In the far Higgsino limit, the chargino and neutralino spectrum
simplifies, and for $M_2\gg|\mu|$,
\begin{equation}
\begin{array}{@{\extracolsep{-0.05in}}ccccc}
m_{\chi_1^0}&\approx& m_{\chi_2^0}&\approx &|\mu|\\
m_{\chi^\pm}&\approx &|\mu|,\\
\label{spectrum}
\end{array}
\end{equation}
A non-zero gaugino component  provides some splitting of the
spectrum, and at tree level,   
\begin{equation}
\begin{array}{@{\extracolsep{-0.05in}}ccccc}
m_{\chi_1^0}&<& m_{\chi^\pm}&<&m_{\chi_1^0}.\\
\end{array}
\end{equation}
The LEP 182 lower bound of $91\gev$ on the chargino mass then imposes a
comparable lower bound on $\mchi$ in the limit of pure Higgsinos.
However, if $\mchi>m_W$, the annihilation channel
$\neut\neut\rightarrow W^+W^-$ is available to neutralinos at their time of 
decoupling in the early universe.  Neutralinos in fact love to
annihilate into $W$ pairs, since this process is not \hbox{p-wave} suppressed,
in contrast to neutralino annihilation into fermion pairs, and so
$\ohsq\sim\langle\sigma_{\rm ann}v\rangle^{-1}$ is greatly reduced
above the $W$ threshold.  Consequently, Higgsinos with masses above $m_W$ (and
below $\sim 1\tev$) are not viable dark matter candidates.  For this
reason, LEP chargino searches are fatally squeezing Higgsino dark
matter.  In this section we examine over how much of the MSSM
parameter space Higgsino dark matter remains viable, and comment on
the fate of Higgsino dark matter when the last LEP runs are complete.

\subsection{Loop Corrections to Ino Masses}
\begin{figure}[thb]
\begin{center}
\vspace*{-0.1in} 
\epsfig{file=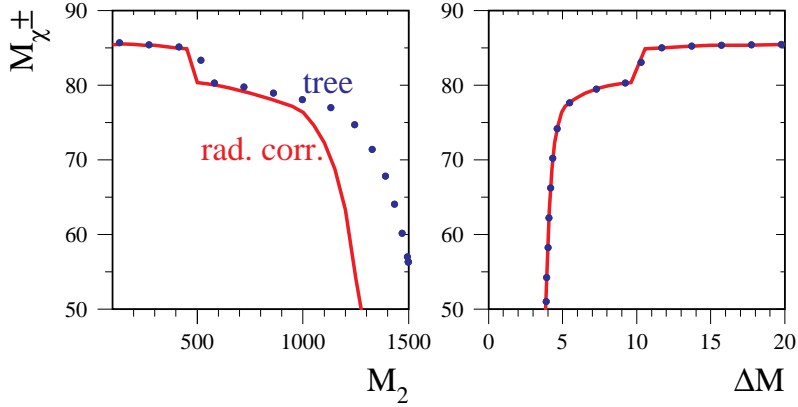,height=2.3in}
\end{center}
\vspace{-0.3in} 
\caption{{\it The experimental limit on $m_{\chi^{\pm}}$ as
    function of $M_2$ and as a function of $\Delta M \equiv
    m_{\chi^{\pm}} - m_{\chi }$, for fixed $m_0 = 200\gev$ and
    $\tan\beta = 2$.  The dotted
    line comes from a tree-level analysis, and the solid line is
    obtained using an ad hoc parameterization of the experimental
    efficiency, in conjunction with the radiatively-corrected mass
    formulae.  }}  
{ \label{fig:mchipm} }
\end{figure}

Constraints on Higgsino dark matter are sensitive to loop corrections
to the neutralino and chargino masses, and this sensitivity appears in
two separate pieces of the analysis.  First, as mentioned in section
\ref{sec:lepbounds}, there is a loophole which appears in the chargino
experimental bounds when the mass of the chargino is close to the mass
of the neutralino, and looking at eqn. \ref{spectrum}, this is
exactly the case in the Higgsino limit, where the chargino and
neutralino masses are both close to $|\mu|$.  The second panel of
Fig.~\ref{fig:mchipm} displays \cite{efgos} the LEP 172 lower bound on
the chargino mass as a function of $\Delta M=\mcha-\mchi$, for $\tb=2$
and $m_0=200\gev$.  Characteristically, the bound on $\mcha$ drops for
small $\Delta M$, and essentially disappears for $\Delta M< 5\gev$.
The first panel in Fig.~\ref{fig:mchipm} displays the chargino bound
as a function of $M_2$, where $|\mu|$ is fixed by the chargino mass.
For larger $M_2$ the neutralino is more pure Higgsino, the masses of
the chargino and neutralino are more degenerate, and the bound on
$\mcha$ drops.

Since the chargino bounds are closely dependent on $\Delta M$, it is
important to consider the 1-loop radiative corrections to the
neutralino and chargino masses in the Higgsino region.  The 1-loop
corrections to the neutralino and chargino masses have been computed
\cite{LTT,PP,PBMZ}, and a comparison of the dotted and solid curves in
Fig.~\ref{fig:mchipm} shows their effect on the chargino bound.  As
expected, at large $M_2$, where $\Delta M$ is small, the effect is
significant, while for $M_2\le400\gev$, there is little effect.  The
overlap of the two curves in the second panel of Fig.~\ref{fig:mchipm}
demonstrates that it is really the quantity $\Delta M$ to which the
chargino bound is sensitive.

The second piece of the analysis which is sensitive to the radiative
corrections is the computation of the Higgsino relic abundance.
Unlike in the gaugino case, the approximate degeneracy of the
neutralino and chargino spectrum (\ref{spectrum}) in the Higgsino
region requires the inclusion of the next-to-lightest and
next-to-next-to-lightest states and their co-annihilations with the
LSP (and self-annihilations) in the calculation of $\ohsq$
\cite{co1,co2}.  That this effect is particularly important is due again
to the p-wave suppression of the annihilation of neutralinos into
fermions.  Expanding the thermally averaged annihilation cross-section
in powers of $(T/\mchi)$,
\begin{equation}
\langle\sigma_{\rm ann}v\rangle_{\ss T}=A \;+ \;B\left(T/\mchi\right) \;+\;\;\ldots,
\end{equation}
one finds the zeroth order piece $A$ suppressed by $m_f^2$.
Neutralinos with masses of interest for dark matter fall out of
chemical equilibrium with the thermal bath when the temperature is
$\sim (1/20-1/25) \times \mchi$, so the effect of \hbox{p-wave}
suppression, then, is an order of magnitude reduction in the
annihilation rate.  By contrast, the co-annihilation processes
\begin{eqnarray*}
\neut_1 \;\cha_1&\longrightarrow& e^+\nu,\;\ldots\\
\neut_1 \; \neut_2&\longrightarrow& f \bar f
\end{eqnarray*}
do not involve the annihilation of identical particles and so do not
exhibit p-wave suppression.  However, the number density of the
heavier scattering states is Boltzmann suppressed at low temperatures,
and so the ratio of co-annihilation to annihilation rates goes as
\begin{equation}
R_{\neut_1 \cha_1}/R_{\neut_1 \; \neut_1}\;\sim\;
(T/\mchi) \;e^{-\Delta M/T}\;\sim\;25\;e^{-25(\Delta M/\mchi)},
\end{equation}
and similarly for $\neut_1 \; \neut_2$ co-annihilation\footnote{ There
  are additional factors which suppress the annihilation rate for
  Higgsinos and increase the ratio $R_{\neut_1
    \cha_1}/R_{\neut_1 \; \neut_1}$.}.  For a degenerate spectrum, this
amounts to better than an order of magnitude increase in the
annihilation rate, whereas if the mass difference $\Delta M/\mchi$ is
as much as 25 percent, this ratio of rates is less than 0.05.  Clearly
there is a tremendous sensitivity of $\ohsq$ to $\Delta M$, and
consequently to the radiative corrections to the chargino and
neutralino masses.

Lastly, we note that the loophole in the chargino bound occurs for
small $\Delta M$, where co-annihilations typically make the Higgsino
relic abundance tiny.  By contrast, the relic density turns out to be
significant only for larger $\Delta M$, where the chargino bounds are
strongest.  It is therefore difficult to wriggle out of the the
chargino mass constraints by appealing to this loophole in the
chargino limits, while still preserving Higgsino dark matter.

\subsection{The Fate of Higgsino Dark Matter}

We now explore the Higgsino dark matter regions which survive the LEP
bounds.  In order to conservatively estimate their area in the
following, we take a large sfermion mass $m_0=1\tev$.  This has both
the effect of minimizing the impact of the Higgs mass constraint by
producing large radiative corrections to the Higgs mass and of
minimizing the contributions to neutralino annihilation through the
neutralino's (small) gaugino component.  We take a large pseudoscalar
mass $m_A=1\tev$, similarly to give a large Higgs tree-level mass and
to minimize the effect of s-channel pseudoscalar annihilation.  We
take $A_t$ at the quasi-fixed point $\sim 2.25\;M_2$.  This is a very
good approximation at low $\tb$, and the small flexibility one has to
vary $A_t$ at larger $\tb$ doesn't substantially impact the allowed
dark matter regions.  Lastly, our working definition of a Higgsino is
that $p^2=\gamma_i^2+\delta_i^2 > 0.81$.  We note that since Higgsino
and gaugino purities add in quadrature to 1, a Higgsino purity of 0.9
(our choice) corresponds to a gaugino purity of 0.44, so our choice is
not too restrictive.

\begin{figure}[ht]
\hspace*{-0.1in}
\vspace*{-1.2in}
\begin{center}
\begin{minipage}{8in}
\hspace*{-0.2in}
\epsfig{file=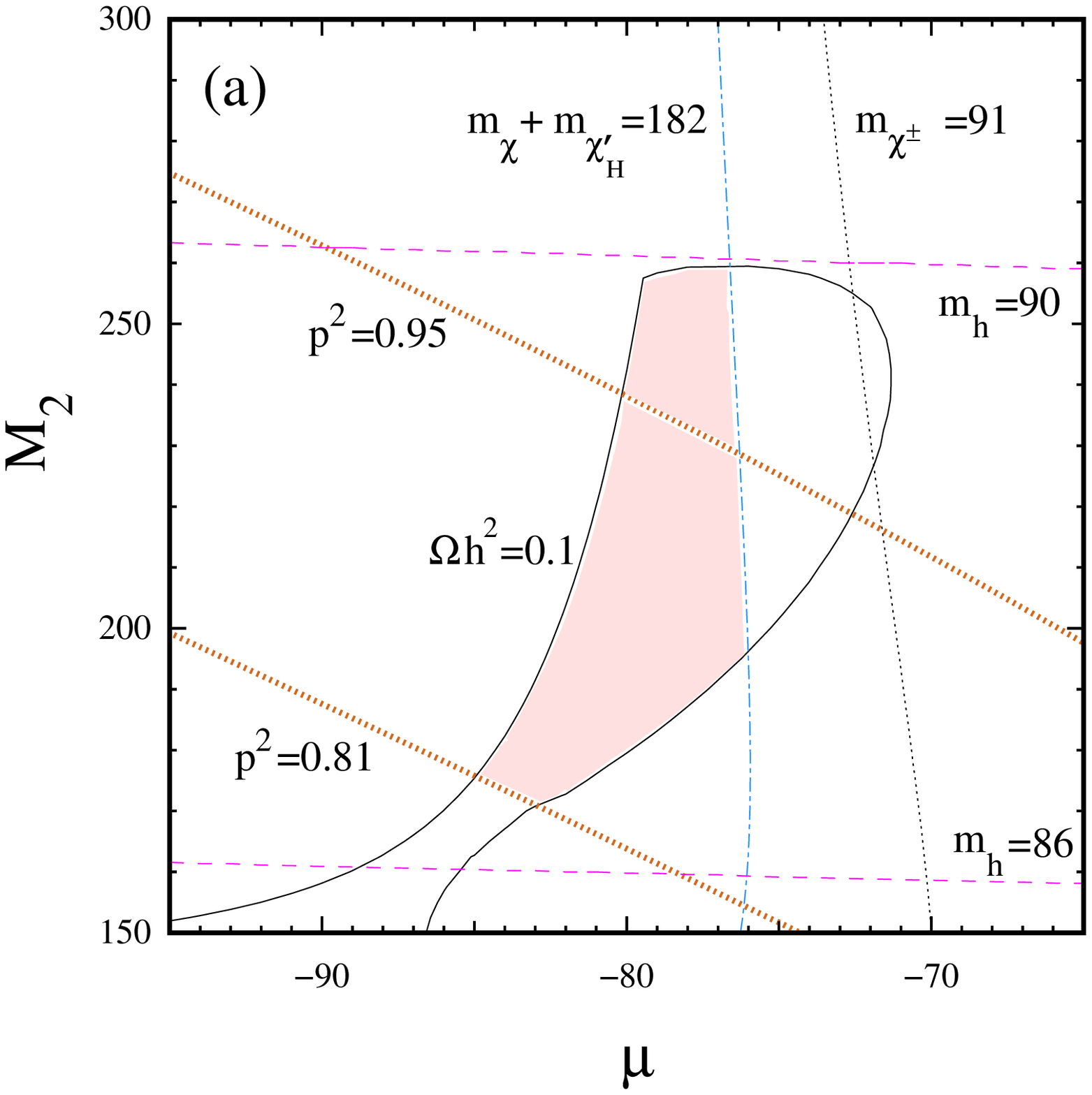,height=3.9in}
\hspace{-1.0in}
\epsfig{file=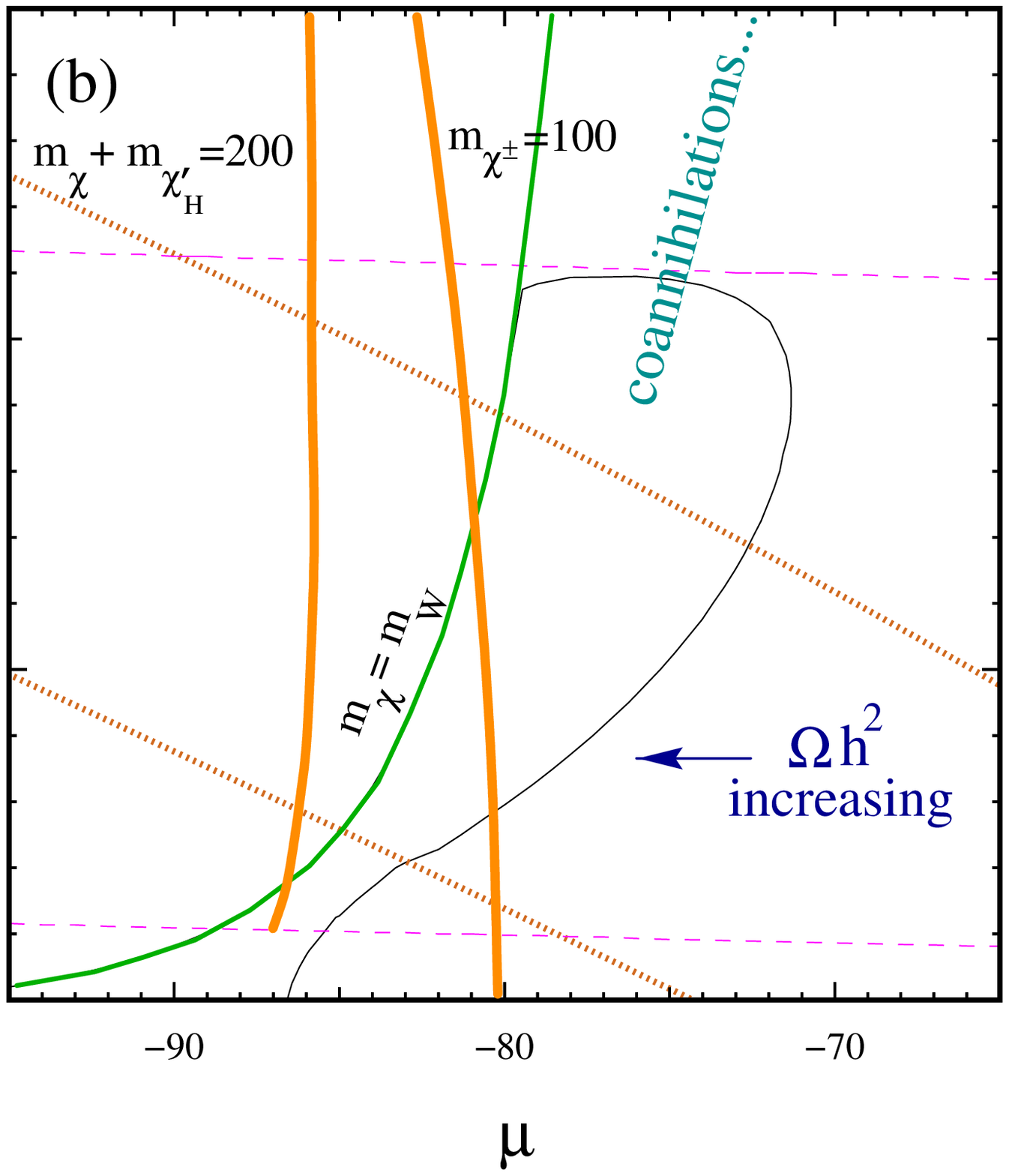,height=3.9in}
\end{minipage}
\end{center}
\vspace{-0.8in} 
\caption{{\it Survey of experimental and cosmological
    constraints in the $\mu, M_2$ plane, focusing on Higgsino dark
    matter for tan$\beta = 2$ and $\mu < 0$ In a) we plot the
    radiatively-corrected contours for $m_{\chi^{\pm}} = 91\gev$, for
    $m_{\chi} + m_{\chi'_H} = 182\gev$, for selected values of $m_h$
    and the Higgsino purity $p$, and for $\ohsq = 0.1$.  The shaded
    regions yield a Higgsino which satisfies the mass and relic
    density constraints described in the text.  In b) we show the
    expected bounds from LEP 200.}}  
{ \label{fig:figsm} }
\end{figure}


In Fig.~\ref{fig:figsm}a we display \cite{efgos} for $\mu<0$ contours
of constant chargino mass $m_{\chi^{\pm}} = 91\gev$, $m_{\chi} +
m_{\chi'_H} = 182\gev$, Higgs mass $m_h$, and Higgsino purity, along
with a contour of constant $\ohsq=0.1$, for $\tb=2$.  We see that the dashed
lines in Fig.~\ref{fig:figsm}a representing the chargino and
associated neutralino mass contours bound one away from small $|\mu|$,
while the Higgs mass limit bounds one away from small $M_2$. The
latter is particularly restrictive at low $\tb$, where the tree-level
Higgs mass is small, and thus where radiative corrections to $m_h$
must be enhanced by taking large stop masses.  The solid contour
contains the region which leads to a significant neutralino relic
density $\ohsq \ge 0.1$, and its limited range in $M_2$ is a result of
co-annihilations.  For larger values of $M_2$, the neutralino is a
purer Higgsino, and the masses of both the lightest chargino and the
next-to-lightest neutralino approach the neutralino mass from above,
enhancing the effect of co-annihilations that deplete the relic
Higgsino abundance.  For larger values of $|\mu|$, the relic density
is suppressed by annihilations into $W$ pairs.  The hashed contours in
Fig.~\ref{fig:figsm}a represent Higgsino purities.  Note the limited
range of $\mu$ for which the mass and relic density constraints are
satisfied.

\begin{figure}[ht]
\vspace*{-1.0in}
\begin{center}
\hspace*{-0.2in}
\epsfig{file=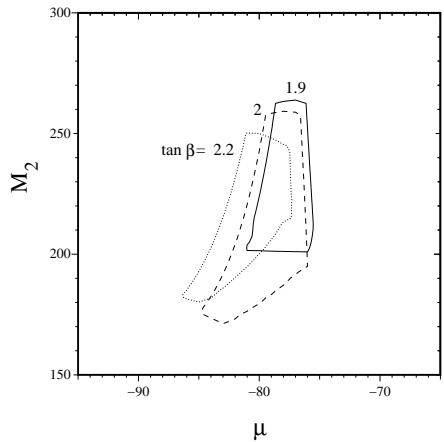,height=3.9in}
\end{center}
\vspace{-0.8in} 
 \caption{{\it The regions of the $\mu, M_2$ plane
     allowed by the constraints shown in the previous figure are shown
     for several different values of tan$\beta$.  There are no
     consistent choices of Higgsino parameters for $\tan\beta < 1.8$
     or $ > 2.5 $ for $\mu < 0$, or for any value of $\tan\beta$ for
     $\mu > 0$.}}  { \label{fig:puddles} }
\end{figure}

The combined effects of the above constraints, corresponding to the
shaded regions of Fig.~\ref{fig:figsm}, are displayed for different
values of tan$\beta$ in Fig.~\ref{fig:puddles}.  We find that there
are no consistent Higgsino dark matter candidates for $\tan\beta \le
1.8$ or $\ge 2.5$ for $\mu < 0$, or for any value of $\tan\beta$ for
$\mu > 0$.  The Higgs mass constraint cuts off the bottom of the
allowed regions at low $\tan\beta$. When $\mu<0$ it becomes a relevant
constraint for $\tan\beta<2.0$ (its effect can be seen in the flat
lower edge of the $\tb=1.9$ contour) and is responsible for the
complete disappearance of the allowed region when $\tan\beta \le 1.8$.
Within the allowed regions displayed, the relic densities generally
increase as $|\mu|$ is increased, until the neutralino mass, whose
minimum value here is $\sim 71\gev$, becomes greater than $m_W$, at
which point the $W^+ W^-$ annihilation channel opens, driving the
relic $\ohsq$ below 0.1.  In Fig.~\ref{fig:figsm}b, we plot the
contour $\mchi=m_W$, which makes evident the drop in $\ohsq$ above the
$W$ pair threshold\footnote{Sub-threshold annihilation of neutralinos,
  not included in this analysis, exclude a further slice of Higgsino
  dark matter on the right-hand side of the $\mchi=m_W$ contour\cite{DNRY}.}  .
In any event, $\ohsq$ is never greater than 0.12 anywhere in the
allowed regions for $\mu<0$.

In Fig.~\ref{fig:figsm}b we show the same plot as
Fig.~\ref{fig:figsm}a, but with an estimate of the bounds that should
be achieved at LEP's eventual run at $\sim 200\gev$ center-of-mass
energy.  We see that the associated neutralino production bound of
$m_{\chi} + m_{\chi'_H} = 200\gev$ excludes entirely the remaining
allowed Higgsino dark matter region at $\tb=2$ for $\mu<0$.  It is
also apparent that at this value of $\tb$, the projected Higgs mass
bound of $\sim 105\gev$, which lies well off the top of the plot, will
play the same role.  And, in fact, the entire set of allowed Higgsino
dark matter regions displayed in Fig.~\ref{fig:puddles} for all $\tb$ 
are excluded by either one of the conditions $m_h>100\gev$ or
$m_{\chi} + m_{\chi'_H} > 200\gev$ alone.  The first of these bounds
should be achieved even if LEP falls somewhat shy of $200\gev$ in its
final run.

\begin{figure}[ht]
\vspace*{-1.0in}
\begin{center}
\hspace*{-0.2in}
\epsfig{file=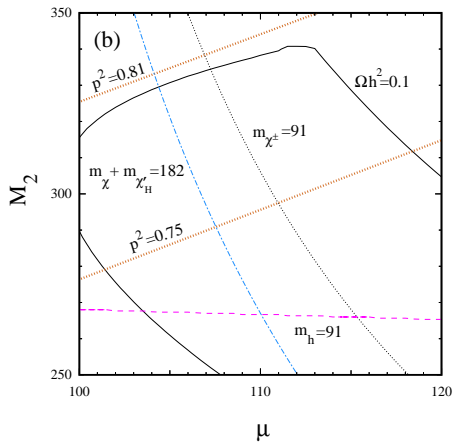,height=3.9in}
\end{center}
\vspace{-0.8in} 
 \caption{{\it Same as Fig.~\ref{fig:figsm}a for $\mu>0$ }}
{ \label{fig:figsmb} }
\end{figure}

In Fig.~\ref{fig:figsmb} we show the equivalent plot to
Fig.~\ref{fig:figsm}a, for $\mu>0$; in this case the Higgsino purities
are lower, and the entire dark matter region falls below the purity
cutoff.  This turns out to be the case for all $\tb$; that is, for all
values of $\tb$, the only regions of parameter space for $\mu>0$ which
have a significant amount of neutralino dark matter $\ohsq>0.1$ have
either a mixed or gaugino-like lightest neutralino.  Lastly, we find the
Higgsino dark matter regions are even more restricted for both
$\mu<0$ and $\mu>0$ if the gaugino masses are related by $M_1=M_2$ 
\cite{Kane}.  For a more detailed
discussion of both the $\mu<0$ and $\mu>0$ cases, see reference
\cite{efgos}.

\section{Summary and Outlook}

We've seen that R-parity conserving SUSY models still provide a good
dark matter candidate.  Neutralinos are of particular cosmological
interest when the lightest neutralino is a gaugino, in particular a
bino, and current experiments are probing into the heart of the
gaugino dark matter region.  Cosmological constraints combine neatly
with experimental bounds, and taken together they exclude all values
of $\tb<2 \;(1.65)$ for $\mu <0 \;(>0)$ in mSUGRA.  By contrast, Higgsino
dark matter is being fatally squeezed by LEP.  Already small,
restricted to an area $\sim10\gev$ wide in $\mu$ and for only $\tb$
between 1.8 and 2.5 for $\mu<0$, the remaining Higgsino dark matter
regions (with masses $<1\tev$) will be finally excluded by either of
the conditions $m_h>100\gev$ or $m_{\chi} + m_{\chi'_H} > 200\gev$,
bounds which should be achieved by LEP 200.

\vspace{0.5in}
\noindent{ {\bf Acknowledgements} } \\
\noindent  
I would like to thank the organizers of the Symposium on Lepton and
Baryon Number Violation for an interesting an enjoyable conference.  I
wish also to gratefully acknowledge John Ellis, Gerardo Ganis, Keith
Olive, and Michael Schmitt, with whom this work was done in
collaboration.  This work was supported in part by DOE grant
DE--FG02--95ER--40896 and in part by the University of Wisconsin
Research Committee with funds granted by the Wisconsin Alumni Research
Foundation.

\end{document}